# INTEGRATION OF MACHINE LEARNING TECHNIQUES TO EVALUATE DYNAMIC CUSTOMER SEGMENTATION ANALYSIS FOR MOBILE CUSTOMERS

Cormac Dullaghan and Eleni Rozaki

School of Computing, National College of Ireland, Dublin, Ireland

## ABSTRACT

*The telecommunications industry is highly competitive, which means that the mobile providers need a business intelligence model that can be used to achieve an optimal level of churners, as well as a minimal level of cost in marketing activities. Machine learning applications can be used to provide guidance on marketing strategies. Furthermore, data mining techniques can be used in the process of customer segmentation. The purpose of this paper is to provide a detailed analysis of the C.5 algorithm, within naive Bayesian modelling for the task of segmenting telecommunication customers behavioural profiling according to their billing and socio-demographic aspects. Results have been experimentally implemented.*

## 1. INTRODUCTION

More mobile customers are using large amounts of data as they watch videos, live streaming programs, and view large numbers of pictures via faster 4G networks. As customers consume larger amounts of data because of the activities in which they engage on their mobile devices, mobile data service revenues for mobile providers also increase. In addition, as more apps for iOS and Android devices are developed, more customers download more apps and use them, which further increases their data usage.[1] Because of increasing mobile data prices, large numbers of subscribers churn from one provider to another in search of better rates. They also churn providers in order to receive benefits for signing up with a new carrier, such as receiving a free or deeply discounted phone. In addition, the lower signup fees associated with prepaid mobile services also encourages customers to churn. The ability of mobile customers to keep their existing mobile numbers through the Wireless local number portability (WLNP) reduces barriers to churning within the industry, which is a major problem for companies in the telecommunications industry.[2] Because of the likelihood of customers to change providers, the deals that telecommunication companies offer may differ based on the needs of individual customers and their wiliness to pay for particular services.

iD Mobile Ireland are the company that is the basis for the work performed in this paper. They are a start-up telecommunications provider in the Republic of Ireland. The company differentiates itself in the competitive Irish market by separating the mobile tariff from the handset. This allows customers the flexibility to enter or leave a 12, 18 or 24 month contract without penalty, and purchase a new handset every three months, should they wish to do so, once the previous handset cost is fully paid off. Additionally as the customer is not tied down into an extended contract where the cost of the phone is subsidised by the tariff price, customers may change their tariff call, text and data allowances every month, to suit their individual needs, allowing them more control over their account charges. They could, for example, increase their call minutes bundle amount for the month of December should they envisage making more calls during this peak holiday period. The company has access to a wide range of data, with the prospect to capture even more data, growing at a rapid rate. The data that the company can access are currently not being used to their full potential as a means of understanding the customers that are served, their sale patterns, the potential fraud risks, and churn patterns. In this paper, the goal is to collect, clean, categorise, and gain insight from a large dataset spanning 16 months of Bill Pay customer account data that contains 26717 rows and 86 columns of attributes. The data also contain an additional 11 columns comprised of formula derived values or classes used to categorise the data. The primary aim of this effort is to better meet customer needs, improve customer satisfaction, developing customer loyalty to the brand as a means of improving customer retention.

The initial step in carrying out this effort was to acquire the relevant data to generate the various reports, using the attributes available, and to cross check the results in the production customer care system as a means of confirming the accuracy of the data. This verification step proved to be very important because as several tables of data where combined, erroneous results occurred. This meant that separate reports had to be created because all of the data could not be in one report due to the database tables not containing the required logic to be joined together or the report data outputted exceeding the current maximum capable by the system, which was 70,000 cells of data.

The primary contribution of this paper is churn prediction improvement through the process of applying some well-known machine learning algorithms to perform customer data segmentation. After experimenting with two different algorithms, decision tree rules models were introduced that showed the segmentation rules within a Bayesian modelling scheme that displayed the significance of the probability of predicting customers buying patterns. [3] The methodology that is suggested in this study indicates how to improve services for VIP customers. However, this methodology includes a tuning parameter that can be manipulated to predict different customer preferences on deals and the probability of a customer switching to a different telecommunication provider. As a result, the machine learning algorithms can be used to predict which customers are likely to switch carriers.[3]

It should also be noted that another important purpose of this project was to provide a way for telecommunication companies to be able to target deals and special programmes or incentives in order to prevent customers from churning. Targeted proactive programs have the potential advantages for telecommunications companies of having lower incentive costs due to the fact that customers are not trained to negotiate for better deals under the threat of churning.[2]

## 2. CUSTOMER RELATIONSHIP MANAGEMENT IN MOBILE COMMUNICATIONS ENVIRONMENT

Large numbers of companies are adopting CRM technology as a way of managing customer relationships. CRM is used to create, maintain and enhance strong relationships with customers and stakeholders.[1] By taking advantage of machine learning techniques with the CRM database, it is possible to find the best customers. This process is known as customer classification.[1]

Machine learning techniques can be used to extract important customer information from a much larger set of data that may be irrelevant for a particular purpose, such as preventing customer churn. Many CRM studies are based on the use of decision trees. Data mining can be used to discover what might otherwise be hidden customer behaviours from large amounts of data. In this regard, the real value of data mining is the ability to transform large amounts of raw data into usable data to address business problems.[1]

The idea is to group and profile customers according to different socio-demographic aspects (age, gender, location, etc.) It is important to notice that studies including these variables allow identifying specific demographic segments of the population to be the target of certain policies and strategies.

However, based on T.Garín-Muñoza, T.Pérez-Amaralb,N.C.Gijónb_R.López 2015 the main complaints of telecommunication customers are based on the following types:

**Reputation of the company:** The likelihood that the company will actually investigate and address a customer complaint.[4]

**Level of expenditure:** The higher the cost on mobile phone services to the customer, the greater the likelihood to complain.[4]

**Overall level of satisfaction:** The overall level of satisfaction of the customer, including an assessment of a problem situation in light of overall satisfaction, and the evolution of a particular problem over time.[5]

**Type of contract:** It is expected that consumers who have post-paid contracts will have a higher likelihood to complain than prepaid customers. The reason for this is because prepaid customers will not incur problems related to billing or breach of offers as they do not receive a monthly bill.[4]

## 2.1. Mining churning behaviours

The variables that are generally used for market segmentation are Demographic, socioeconomic, and geographic characteristics of the customers. A very useful technique for behavioural-based data mining method in the RFM analysis, which involves the extraction of customer profiles by using a few criteria, which reduces the complexity of analysis.[5] In RFM analysis, customer data are classified by Recency (R), Frequency (F) and Monetary (M) variables. It has been noted that RFM enables the practitioners to observe customer behaviour, as well as to segment customers in order to determine immediate customer value.[5]

It should also be noted that using decision rules algorithms for the purpose of customer segmentation may result in an efficient evaluation of a segmentation plan. Decision trees can be identified into sets of if-then rules, which means that they can be used to solve a variety of problems, such as customer segmentation and customer churn prediction. In fact, many researchers have used this method to study customer segmentation. A customer satisfaction survey can be used to construct a customer segmentation system based on demographic variables and even customer reviews.[1]

Researchers have provided ideas about modelling customer satisfaction using unstructured data with a Bayesian approach. They explain that the transformation of unstructured data taken from customer's reviews into a semi-structured form associated with each aspect reflecting the frequency counts for positive, negative, and neutral sentiments. One assumption of this model is that the rating of each aspect is based on a particular combination of the positive, neutral and negative sentiments of that particular aspect. The result is that the overall aspect rating depends upon how many times an aspect has been associated with positive, neutral and negative sentiments in a single customer review. Furthermore, there is also an overall rating that is assigned to each review by the contributor.[6]

## 2.2. Applications of Data Mining Techniques in Telecom Churn Prediction

The data regarding user communication characteristics consists of Call Data Record information. The telephone usage habits of customers are indicated by the start time, the number of users to make the call, the sum, duration, call type, call variation information, and other statistical data. As fraudulent behaviour may have a fixed behaviour pattern, most fraudulent behaviour can be found by examining the CDR data. It is typically the case that normal users sometimes delay payment because of special reasons or habitual delays. However, these types of delays are different from fraudulent behaviour as they generally have a fixed behaviour patterns. In order to identify late paying users, the late payment behaviour must be compared with customer payment records.[7] Fraudulent behaviour can result in telecom providers suffering heavy short-term financial losses. Late payments that are due to non-fraudulent behaviour may not immediately cause significant financial losses, but they may result in other types of losses such as cash flow reductions, increased labour costs related to debt collected customer churn.[7]

Based on the information that has been reviewed, possible variables for modelling the decision tree were selected.[8] The creation of a test to determine proper costs, as well as to prevent cross subsidisation is difficult in the telecommunications industry. The difficulty arises from the pervasive common and joint costs that arise within the industry. Telecommunications companies typically group multiple services rather than providing individual services that are unrelated to each other. This means that investment costs are common to multiple services.[8] Among them, the most significant cost variables for customers evaluation that have higher contribution to predict the churn are selected.

## 3. CUSTOMER BEHAVIOUR PATTERN ANALYSIS AND DERIVED ATTRIBUTES

One of the issues that arose in this study involved transforming the raw data. An effort was made to collect, clean, categorise, and gain insight from a large dataset that contained 26717 instances and 86 attributes with a further 11 columns comprised of formula derived values or classes to categorise the data. In order to deal with this issue, the derived attributes were performed based on the needs of the learning machine scheme.[6]

It is important to understand that derived attributes are new variables that are based on original variables. The most effective derived variables are those that represent something in the real world, such as customer behaviour. There are general classes of derived variables, such as total values, average values, and ratios. In this study, the derived variable of the average value over the last six months is used. In addition, the ratio between the average value over the last three months and the average value over all months is used.[9]

### 3.1. Customer Demographic Profiles

Customer demographic profiles are a grouping of a demographic or market segment that contains likely consumer behaviour. This information typically includes age, location and gender, among other demographic variables. The information of gender and counties are available and selected as two new features in this study.[10]

### 3.1.1 Customer Age Group

The extracted attribute of customer age groups categorises the customer's age into five groups, 0-14, 15-24, 25-44, 45-64 and 65+. These groups were chosen based on the Irish Census groupings to allow for future comparisons to be made. First it performs an error check to ensure the age column is a number, then it checks which group that number falls within and returns that group value. The decision tree classifier facilitates the process of classification in regards to the customers age.

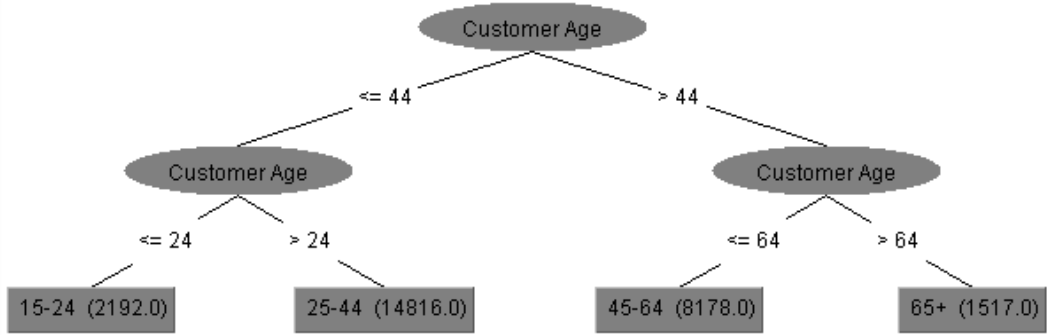

Figure 1. Decision tree of Customer age

### 3.1.2 Customer Location County

This variable is the approximate county location of the customers obtained from the county name in the Customer Bill Address column. Because the data are manually entered at point of sale, they are not consistent and may not always contain the county name. However, Eircodes, which is Ireland's Post Code, has become mandatory as part of the sales process, which means that precise location categorisation can occur in the future.

```
=== Confusion Matrix ===

     a    b    c    d    e    f    g    h    i    j    k    l    m    n    o    p    q    r    s    t    u    v    w    x    y    z   aa   ab   ac   ad   <-- classified as
  2608    1    0    0    0    1    0    2    3    1    2    0    0    2    1    1    0    0    3    0    0    1    1    0    0    0    2    0    0    0 |   a = Dublin
    19  437    0    0    0    0    0    0    0    0    0    0    0    0    0    1    0    0    0    0    0    0    0    0    0    0    0    0    0    1 |   b = Galway
    10    0  122    0    0    0    0    0    0    0    0    0    0    0    0    0    0    0    0    0    0    0    0    0    0    0    0    0    0    1 |   c = Donegal
    37    0    0  652    0    2    0    0    0    0    1    0    0    0    0    0    0    5    0    0    1    0    0    0    0    0    0    0    0    1 |   d = Cork
     2    0    0    0   85    0    0    0    0    0    0    0    0    0    0    0    0    0    0    0    0    0    0    0    0    0    0    0    0    1 |   e = Carlow
   188    2    0    0    0  419    0    6    0    0    0    0    0    0    1    0    0    0    0    0    0    0    0    0    0    0    0    0    0    1 |   f = #N/A
     5    0    0    0    0    0  111    0    0    0    0    0    0    0    0    0    0    0    0    0    0    0    0    0    0    0    0    0    0    1 |   g = Leitrim
    34    2    0    0    0    1    0    3    1    0    3    1    0    3    0    2   10    4    0    0    2    4    0    2    0    3    0    2    0    1 |   h = Down
    25    0    0    0    0    0    0  434    0    0    0    0    0    0    0    0    0    0    0    0    0    0    0    0    0    0    0    0    0    1 |   i = Meath
     7    0    0    0    0    0    0    0    1   89    0    0    0    0    0    0    0    0    0    0    0    0    0    0    0    0    0    0    0    1 |   j = Longford
     8    0    0    0    0    0    0    0    0    0  342    0    0    0    0    0    0    0    0    0    0    0    0    0    0    0    0    0    0    1 |   k = Limerick
     9    4    3    0    0    1    1    0    3    1    0    1    2    4    1    1    2    4    0    3    1    3    0    1    3   18    0    0    0    1 |   l = Derry
    10    0    0    0    0    0    0    0    0    0    0    0  350    0    0    0    0    0    0    0    0    0    0    0    0    0    0    0    0    1 |   m = Wexford
    17    0    0    0    0    0    0    0    0    0    0    0    0  144    0    0    0    0    0    0    0    0    0    0    0    0    0    0    0    1 |   n = Laois
     5    0    0    0    0    0    0    0    0    0    0    0    0    0  183    0    0    0    1    0    0    0    0    0    0    0    0    0    0    1 |   o = Kilkenny
    24    0    0    0    0    0    0    2    0    0    0    0    0    0    0  313    0    0    0    0    0    0    0    0    0    0    0    0    0    1 |   p = Kildare
     7    1    0    0    0    0    0    0    0    0    0    0    1    0    0    0  156    0    0    0    0    0    0    0    0    0    0    0    0    1 |   q = Roscommon
     9    0    0    0    0    0    0    0    0    0    0    0    0    0    0    0    0  473    0    0    0    0    0    0    0    0    0    0    0    1 |   r = Waterford
    10    0    0    0    1    0    0    0    0    0    0    0    0    0    0    0    0    0  134    0    0    0    0    0    0    0    0    0    0    1 |   s = Louth
     9    0    0    0    0    0    0    0    0    0    0    0    0    0    0    0    0    0    1    0  156    0    0    0    0    0    0    0    0    1 |   t = Mayo
    15    0    0    0    0    0    0    0    0    0    0    0    0    0    0    0    0    0    0    0    0  194    0    0    0    0    0    0    0    1 |   u = Tipperary
     4    0    0    0    0    0    0    0    0    0    0    0    0    0    0    0    0    0    0    0   36    0  178    0    0    0    0    0    0    1 |   v = Monaghan
    15    0    0    0    0    0    0    0    0    0    0    0    0    0    0    0    0    0    0    0    0    0    0  179    0    0    0    0    0    1 |   w = Wicklow
     1    0    0    0    0    0    0    0    0    0    0    0    0    0    0    0    0    0    0    0    0    0    0    0   99    0    0    0    0    1 |   x = Sligo
    11    0    0    0    0    0    0    0    0    0    0    0    0    0    0    0    0    0    0    0    0    0    2    0  195    0    0    0    0    1 |   y = Kerry
     3    0    0    0    0    0    0    0    0    0    0    0    3    0    0    0    0    0    0    0    0    0    0    0   46    0    0    0    0    1 |   z = Offaly
    25   21    0    1    0    0    1    0    1    0    3    0    0    1    1    1    0    2    0   18    0    0    1    1    0  119    0    0    0    1 |  aa = Clare
     2    0    0    0    0    0    0    0    0    0    0    0    0    0    1    0    0    0    0    0    0    0    0    0    0    0   63    0    0    1 |  ab = Cavan
     1    0    0    0    0    0    0    0    0    0    0    0    0    0    0    0    0    0    0    0    0    0    0    0    0    0    0    0    0    1 |  ac = Armagh
     1    0    0    0    0    0    0    0    0    0    0    0    0    0    0    0    0    0    0    0    0    0    0    0    0    0    0    0    0    1 |  ad = Tyrone
```

Figure 2. Confusion Matrix from Location County Model

The diagonal of the confusion matrix shows the different costs for the locations classification, all of which are positive values. The use of this type of confusion matrix is to create classifiers that minimise the prediction cost as opposed to the prediction error that occurs in customers' segmentation based on the county. Moreover, the locations with the higher prediction errors in cost benefits shown in Dublin and Cork as well as in Waterford and Meath.

## 3.2. Sales information about Day & Time

The variable for sales information about day & time contains the types of service packages, credit controller indicators, and the first date of using the services. The variable also contains the creation date and time, the bill frequency, the account balance, equipment rents, payment types, and contract duration.[10]

### 3.2.1. Customer Length of Service

This variable contains the total number of days a customer account has been in service. The variable was created by checking the Customer Network Status, and if inactive, minuses the Subscriber inactive Date from the Customer Activation Date. If no inactive date was present, the variable was created by subtracting the Customer Activation Date from today's date to find the current length of service in days.

### 3.2.2. Service Sale Date

The service sales date attribute was created from the value of the sale date column with the format of DD/MM/YYY and converted it to a day in the week. For example, 28/07/2015 was converted to Tuesday.

### 3.2.3. Sale Time of Day

This variable contained the sales time in the 24 hour format HH:MM:SS and categorised into 4 types. If the hour was less than 6, it was defined as Night. If the hour was less than 12, it was defined as Morning. If the hour was less than 17, it was defined as Afternoon. Otherwise, the time of day was defined as Evening. For example, 10:17:55 was categorised as Morning.

## 3.3. Customer account information about bills and payments

This variable contains the billing information for each customer and service for a certain number of years. The last 16-months of Bill Pay customer account data are available and only used in our prediction system. However, different customers may have had different bill occurrences depending on when they joined the mobile network and their length of service. In this regard, the durations of customer bills might have been different.[10]

### 3.3.1. Total Invoice Amount Excluding Brought Forward

This variable contains the Total Invoice Amount Excluding Brought Forward in order to add in every month's invoice amount minus any amount unpaid and brought forward from the previous month. By performing this calculation, it was possible to find the total revenue generated by each customer's account.

### 3.3.2. Total Number of Invoices

The attribute of Total Number of Invoices is to count the total number of invoices a customer has received. It required an additional row be created above the attribute names, with the string "Invoice" being placed above each column of data where the months' invoice amount is. This is due to the data layout having the balance brought forward attributes between each month's invoice amount and the Count function only being able to select a single range of columns.

### 3.3.3. Average Invoice Amount Excluding Brought Forward

The attribute of Average Invoice Amount Excluding Brought Forward checks to see if the value for Total Invoice Amount Excluding Brought Forward is zero, and if so returns zero, otherwise divide it by the value in Total Number of Invoices to find an average invoice amount.

### 3.3.4. Total Paid Amount of Invoices

The Total Paid Amount of Invoices attribute takes a count of every paid invoice. As the attribute data is located in columns beside each other, a single range can be defined as paid, unpaid and decline status of the invoices. N is a set of Total Paid Amount of Invoices attributes: $N = \{P_{ji,}\ DCl_{ji,}\ U_{ji,}\}$

# 4. CUSTOMER SEGMENTATION MODEL

A set of rules can be generated for checking customer willingness to pay based on average invoice amount. The rule that was created was that if the average invoice amount is less than €15, then define as "Low Spender".  If the average invoice amount is between €15 and €29, then define as "Average Spender" If the average invoice amount is between €29 and €50, then define as "Above Average Spender".  If the average invoice amount is between €50 and €70, then "High Spender". If the average invoice amount is above €70 then "Very High Spender". If none of these apply, then "Investigate" is returned. The $ClassA_{ji}$  presents the low spender status, $ClassB_{ji}$ presents the Average Spender status, $ClassC_{ji}$ presents the Above Average Spender, $ClassD_{ji}$ presents the High Spender, $ClassE_{ji}$ presents the Very High Spender status.

## 4.1. Learning algorithm: Seeking the customer profile rules set

This algorithm categorises each customer account into four different classes based on their invoices being paid and spender status. If the Total Paid Amount of Invoices is greater than or equal to one less of the Total Number of Invoices, and if their Spender Status is either a Low Spender or an Average Spender, then they are defined as Standard. If the spender status is either a High Spender or an Above Average Spender, then they are defined as Premium. Alternatively, if the spender status is a Very High Spender, then they are defined as VIP. If the Total Paid Amount of Invoices is less than one less of the Total Number of Invoices, then it is defined as Unpaid Invoice to highlight that a customer account is not paid up to date.

**Inputs:**

$N$ is a set of Total Paid Amount of Invoices attributes: $N = \{P_{ji}, DCl_{ji}, U_{jk,}\}$

$F$ is a set of *Customer Length of Service*: $F = \{(Cf, Csc, Ctr, Cfr,)\}$ *Rescaling of range of time for each attribute in G*

$E$ is a set of data related to Spender Status generated from decision tree algorithm:

$E = \{(ClassA_{ji}, ClassB_{ji}, CLassC_{ji}, CLassD_{ji}, CLassE_{ji})\}$

$G$ represents a set of customer values for each attribute in F.

$G = \{(D_{vip}, D_{premium}, D_{standard}, D_{lowspender}, D_{unpaid})\}$ - Rescaling of customers deals regarding the network costs

**Outputs:**

Get values for N and F, the size of the data set and the pattern, respectively.

Set $C_i$, the starting point of each customer for attempted match, to $E$

For each point of $C_i$ in N do If Ci =F/N$_{(m+(i-1)}$ then

    Set the value of Spender Status E to *ClassA_{ji}*, OR *ClassB_{ji}*,

        While account status $N=\{P_{ti}\}$ AND $F=\{(C_{ti})$ OR $(C_{sci})\}$

        do Standard

  For each point of sales $P_i$ in N do If Ci =F/N$_{(m+(i-1)}$ then

  Set the value of Spender status $\{(ClassA_{ji}, ClassB_{ji}, CLassC_{ji}, CLassD_{ji}, CLassE_{ji})\}$ to $CLassC_{ji}$ OR $ClassD_{ji}$,

        While account status $N=P_{ti}$ AND $F=\{(C_{tri})$ OR $(C_{fri})\}$

        do Premium

  Set the value of Spender status $\{(ClassA_{ji}, ClassB_{ji}, CLassC_{ji}, CLassD_{ji}, CLassE_{ji})\}$ to $ClassE_{ji}$,

        While account status $N=P_{ti}$

        do VIP

Else check for "Unpaid Status"

Figure 3. Customer Segmentation algorithm

# 5. EXPERIMENTAL RESULTS AND DISCUSSION

In order to validate the various test option efficiencies, a collection of results were gathered from repeatedly running the same classifier test. The test data must be different to the training data to ensure accurate test results. As all of the data used in this study was from a single dataset, a separate test set file was not used. However, cross-validation and percentage split test options were used. Firstly, in order to select the parameters of the model, the data were divided into training and testing sets. The training data was used for parameter estimation, while the test set was used for evaluation of the methodology.[11] To ensure an accurately different result each of the ten times, a new random seed was provided in the options for the percentage split each time. The Sample Mean was calculated by averaging all ten results, and the Sample Standard Deviation was used to measure the range of the numbers. After all percentage split tests were completed, cross validation was used and set to 10 folds. [12] This meant that the entire dataset was divided into 10 pieces with each used in turn for training. Then, an eleventh test was performed using the entire dataset as the test data. Once completed, the final result was displayed.

## 5.1. Experiment I -Customers segmentation using decision tree classification rules

In this study, the data classification was used to determine the likelihood of a result from creating a decision tree. The attributes of an instance were used against the decision tree to determine the likelihood of the result by progressing through each step until the final decision.  It should be noted that this is one of the most popular and widely used classification techniques.

First, classifying a customer's VIP status is explained. Because of the large number of attributes in the dataset, the C.5 algorithm was originally not available. After removing several derived attributes that were deemed not to be relevant for this classification, the C.5 algorithm with 86 attributes was available for use.

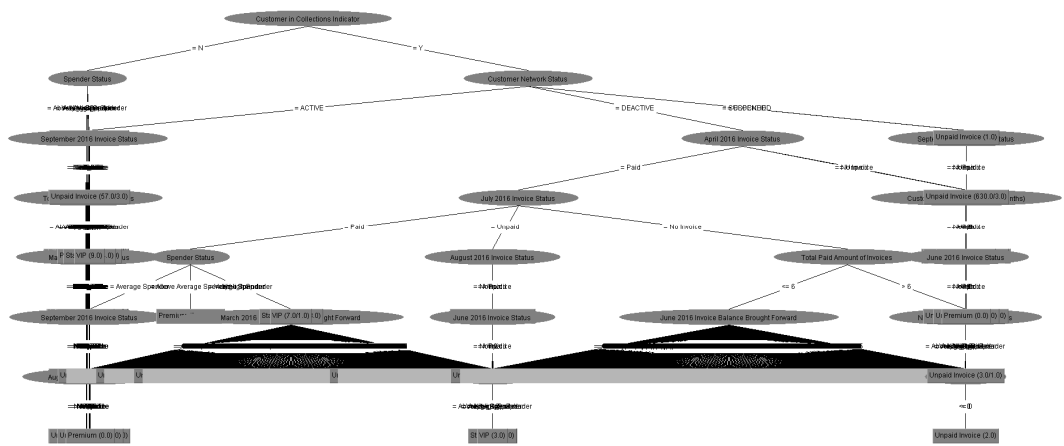

Figure 4. Customer Segmentation decision rules

The decision tree algorithm extracted the rules for customer profiling and evaluation. The segmentation decision rules were based on spender status and the payment behaviours, such as the total amount of paid and unpaid invoices and the customer preferences in regards to the deals offered by the provider. The size of the decision tree was 9124 and had 9062 leaves. The algorithm resulted in classification of the customer profiles into four categories, such as standard, unpaid invoice, premium and VIP status. The rules extracted from the tree shows the most reliable customer, their spending status, frequency of purchases during date and time, how to track unpaid invoices and block accounts and, whether they were VIP customers.

The VIP Status could be deemed to be a successful classifier because of the high accuracy of correctly classified instances. The decision tree algorithm showed additional rules about the VIP churners and their length of time, the services that have received, whether they were more likely to purchase a deal in regards of time and day, and the most popular addresses at which VIP customers were located.

After completing all eleven tests, the Sample Mean of Correctly Classified Instances was 97.70315% for Percentage Split, while Cross-validation 10 Folds showed Correctly Classified Instances of 97.6669%. Given the very slight 0.03625% difference in correctly classified instances, along with the reduced time to run a single cross-validation 10 folds' test, the most efficient test method appeared to be cross validation. [12]

## 5.2. Experiment 2- Bayesian modelling of customer profiles

In this study, the data set was analysed using the Naïve Bayesian modelling. A machine learning technique based on Naïve Bayes model assumes the presence of a particular feature in a customer profile class that is unrelated to the presence of any other feature.

As has been shown, the Bayesian network provided estimations about the utility of every possible attribute value in the spender status domain. In order to use these estimations to elicit customer preferences, a notion was needed regarding the relative importance of the attributes relative to each other. In the approach used in this study, the importance of an attribute depended on three factors: Customer demography, Customer Length of Service Sale Day & Time of Day and financial considerations. [13]

```
Time taken to build model: 0.16 seconds

=== Evaluation on test split ===

Time taken to test model on training split: 0.48 seconds

=== Summary ===

Correctly Classified Instances        7964               87.7189 %
Incorrectly Classified Instances      1115               12.2811 %
Kappa statistic                          0.7882
Mean absolute error                      0.0615
Root mean squared error                  0.2422
Relative absolute error                 20.2438 %
Root relative squared error             62.0806 %
Total Number of Instances             9079

=== Detailed Accuracy By Class ===

         TP Rate   FP Rate   Precision   Recall   F-Measure   MCC     ROC Area   PRC Area   Class
         0.980     0.210     0.844       0.980    0.907       0.793   0.960      0.961      Standard
         0.780     0.002     0.990       0.780    0.873       0.853   0.990      0.976      Unpaid Invoice
         0.771     0.029     0.893       0.771    0.827       0.781   0.964      0.910      Premium
         0.011     0.002     0.059       0.011    0.018       0.021   0.794      0.030      VIP
Weighted Avg.  0.877  0.120  0.879       0.877    0.871       0.794   0.966      0.942

=== Confusion Matrix ===

   a     b     c    d   <-- classified as
 4778     9    85    4 |   a = Standard
  323  1494    86   12 |   b = Unpaid Invoice
  499     4  1691    0 |   c = Premium
   59     2    32    1 |   d = VIP
```

Figure 5. Naïve Bayesian profile classification results

Through the use of semi-structured data, it is proposed that a Bayesian approach be used to model the overall customer preferences in terms of the aspects identified from the consumption and the billing behaviour associated with a customer's willingness to pay. This Bayesian model considered the overall rating of each deal as a weight and sum of the probability to select the individual offers. This model allowed for the determination of an estimation of the tendencies for each deal aspect from each customer's perspective.[6]

The use of the Naïve Bayes model provided a good, but still less accurate result than the C.5 model, of 87.7189% Correctly Classified Instances. The VIP type in the VIP Status class is where the precision is weakest and thus caused the reduced overall accuracy.

One might consider a point in the customer preferences where the model shows a high precision and accuracy related to the "premium offer" as evidence within the naïve Bayesian network leading to the posteriori probability-distributions shown in Fig.5. Evaluating the Influences as have been explained, the learning algorithm used in this study showed that the customer tends to choose a premium offer due to the additional need for cell services and mobile internet connectivity.[13]

The Naïve Bayes results showed that the probability of selecting the premium offer was higher for the customers. However, there was also a very low independence among the preferences related to the "standard deal" offered by the mobile provider. Based on these results, churn management models should not only identify the customers that are most likely to leave the current service provider but also identity the customers that are most likely to respond positively to the right retention deal.[13]

Furthermore, based on the results that were obtained, the "standard deal" needs to be reconsidered and reviewed in relations to customer's need and customer willingness to pay. In addition, a significant probability of fraud detection was found to be classified in the "unpaid

Status" customer profile shown on the Bayesian results that need to be further investigated by the network providers.

# 6. PERFORMANCE EVALUATION RESULTS

Three types of variables were used to develop the two machine learning models that can be tailored according to the customer needs and customer's wiliness to pay. In order to actually implement the plan, more retention efforts should be given to the potential churners who are most likely to react positively.[13]

In this study, the prediction accuracy of each model for each data set in regards to the ROC curve, precision, recall and RPC area were examined.  In addition, the predictions of models built and evaluated on more transactions in the training data, and models built and evaluated on customer's deals were also examined. Furthermore, in addition to the True Churn and False Churn illustrated in the ROC Curve for decision tree models, the results of the Naïve Bayesian classification. PRC, and the overall accuracy were examined, and the results are shown in figures 6&7.

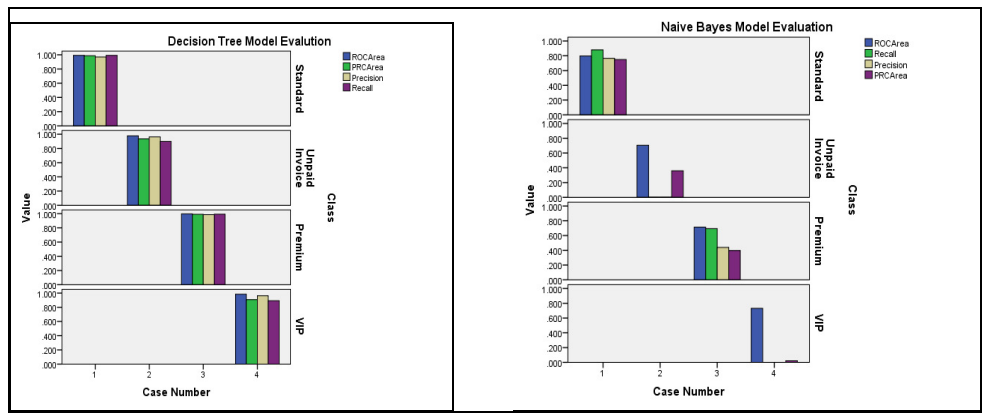

Figure 6. Decision Tree          Figure 7. Naïve Bayes

Figures 6 and 7 show the performance of the model achieved by each machine learning technique and the maximum accuracy that was archived by the decision tree model in all the deals.[15] It should be noted that the main purpose of the Bayesian network was to derive utility estimations for attribute Values from which customers' preferences were derived. In this regard, the churn probability in different deals shown in the Naïve Bayes results was based on the assumption that values that were more likely to satisfy the customer's needs were also more useful.[14]

# 7. CONCLUSIONS

The Use of machine learning techniques with a dataset built upon customer data generated from the Telecommunications Industry made it possible to test various classifiers for categorising a Customer's Age Group, VIP Status, Spend Status and Customer Length of Service.

All of the features used for the churn prediction in this study were either demographic billing, or usage features. The goal was to gain an understanding about the importance of these three types of deals in churn prediction.  The conclusion that was reached was that the billing and usage features had a very high importance for a customer segmentation scheme.[2]

Finally, the two ways in which to evaluate customers' segmentation were equally important for predicting churn. The number assigned to a category of a spender showed the importance of a deal in churn prediction. The result of the decision tree algorithm and preference given to the features are a bit different from the Bayesian model. However, usage and billing features were still of primary importance, while demographics may also affect the churn prediction.[2] In the future, research is planned for mining further churning behaviours and developing retention strategies.

# REFERENCES

[1] L. Luan and H. Shu, "Integration of data mining techniques to evaluate promotion for mobile customers' data traffic in data plan," 2016 13th International Conference on Service Systems and Service Management (ICSSSM), 2016.

[2] A. A. Khan, S. Jamwal, and M. Sepehri, "Applying Data Mining to Customer Churn Prediction in an Internet Service Provider," International Journal of Computer Applications, vol. 9, no. 7, pp. 8–14.

[3] A. Keramati, R. Jafari-Marandi, M. Aliannejadi, I. Ahmadian, M. Mozaffari, and U. Abbasi, "Improved churn prediction in telecommunication industry using data mining techniques," Applied Soft Computing, vol. 24, pp. 994–1012.

[4] T. Garín-Muñoz, T. Pérez-Amaral, C. Gijón, and R. López, "Consumer complaint behaviour in telecommunications: The case of mobile phone users in Spain," Telecommunications Policy, vol. 40, no. 8, pp. 804–820.

[5] A. C. B. Dursun and M. Caber, "Using data mining techniques for profiling profitable hotel customers: An application of RFM analysis," Tourism Management Perspectives, vol. 18, pp. 153–160, 2016.

[6] M. Farhadloo, R. A. Patterson, and E. Rolland, "Modeling customer satisfaction from unstructured data using a Bayesian approach," Decision Support Systems, vol. 90, pp. 1–11.

[7] C.-H. Chen, R.-D. Chiang, T.-F. Wu, and H.-C. Chu, "A combined mining-based framework for predicting telecommunications customer payment behaviors," Expert Systems with Applications, vol. 40, no. 16, pp. 6561–6569.

[8] E. Rozaki, "Financial Predictions Using Cost Sensitive Neural Networks for Multi-Class Learning," Advanced Engineering Forum, vol. 16, pp. 104–116.

[9] V. Umayaparvathi and K. Iyakutti, "Applications of Data Mining Techniques in Telecom Churn Prediction," International Journal of Computer Applications, vol. 42, no. 20, pp. 5–9.

[10] B. Huang, M. T. Kechadi, and B. Buckley, "Customer churn prediction in telecommunications," Expert Systems with Applications, vol. 39, no. 1, pp. 1414–1425.

[11] Brett Lantz, "Machine Learning with R" Packt Publishing Ltd. Second edition July 2015.

[12] I.H. Witten & E. Frank, "Data Mining , practical machine learning tools and techniques," Morgan Kaufmann publishers, 2005.

[13] S. Radde and B. Freitag, "Using bayesian networks to infer product rankings from user needs," UMAP 2010 Workshop on Intelligent Techniques for Web Personalization and Recommender Systems.

[14] H. Lee, Y. Lee, H. Cho, K. Im, and Y. S. Kim, "Mining churning behaviors and developing retention strategies based on a partial least squares (PLS) model," Decision Support Systems, vol. 52, no. 1, pp. 207–216.

[15] Y. Huang and T. Kechadi, "An effective hybrid learning system for telecommunication churn prediction," Expert Systems with Applications, vol. 40, no. 14, pp. 5635–5647.